 \documentclass[final,5p,times,twocolumn]{elsarticle}

\usepackage{amssymb}

\journal{Optics Communications}

\usepackage{color}

\begin{document}

\begin{frontmatter}



\title{Characteristics and stability of soliton crystals in optical fibres for the purpose of optical frequency comb generation}


\author[AIP]{M. Zajnulina}
\author{M. B\"ohm}
\author[AIP]{D. Bodenm\"uller}
\author[ASTON]{K. Blow}
\author[AIP]{J. M. Chavez Boggio}
\author[ARG]{A. A. Rieznik}
\author[AIP]{M. M. Roth}

\address[AIP]{innoFSPEC Potsdam, Leibniz Institute for Astrophysics, An der Sternwarte 16, 14482 Potsdam, Germany}
\address[ASTON]{Aston Institute of Photonic Technologies, Aston Triangle, Birmingham B4 7ET, United Kingdom}
\address[ARG]{Instituto Tecnologico de Buenos Aires and CONICET, Buenos Aires, Argentina}

\begin{abstract}
We study the properties of a soliton crystal, an bound state of several optical pulses that propagate with a fixed temporal separation through the optical fibres of the proposed approach for generation of optical frequency combs (OFC) for astronomical spectrograph calibration. This approach - also being suitable for subpicosecond pulse generation for other applications - consists of a conventional single-mode fibre and a suitably pumped Erbium-doped fibre. Two continuous-wave lasers are used as light source. The soliton crystal arises out of the initial deeply modulated laser field at low input powers; for higher input powers, it dissolves into free solitons. We study the soliton crystal build-up in the first fibre stage with respect to different fibre parameters (group-velocity dispersion, nonlinearity, and optical losses) and to the light source characteristics (laser frequency separation and intensity difference). We show that the soliton crystal can be described by two quantities, its fundamental frequency and the laser power-threshold at which the crystal dissolves into free solitons. The soliton crystal exhibits features of a linear and nonlinear optical pattern at the same time and is insensitive to the initial laser power fluctuations. We perform our studies using the numerical technique called Soliton Radiation Beat Analysis.

\end{abstract}


\end{frontmatter}


\section{Introduction}
\label{sec:intro}
Optical frequency combs (OFCs) are discrete optical spectra with lines that are phase-locked
and have an equidistant spacing as well as nearly equal intensities over a broad spectral range \cite{HolzwarthUdem00,CundiffYen03}. Since their discovery in mode-locked Ti:Sapphire lasers in the 90's, they have been observed in various nonlinear optical media such as semiconductor micro-resonators \cite{DelHayeSchliesser07} and fibre-laser cavities \cite{SteinmetzWilken}. OFCs show a wide range of application potential in telecommunications for the generation of high-repetition rate picosecond-pulses for ultra-high capacity transmission systems based on optical time-division multiplexing \cite{DianovMamyshev89, DudleyGutty01, PitoisFatome02, FinotFatome07, FortierKibler08, FatomePitois10, MansouriFatome11}, spectroscopy \cite{HolzwarthUdem00,ThorpeBals}, metrology, frequency synthesis, and optical clocking \cite{YeCun}. 

In our group, we focus on the deployment of OFCs for the purpose of the astronomical spectrograph calibration. The OFCs generated in mode-locked lasers have been proposed and already successfully tested as calibration sources for high-resolution astronomical spectrographs used for the search for exoplanets or the measurement of the time-variation of the fundamental constants. The spectral line spacings they provide reach up to 50 GHz which is achieved by filtering the mode-locked laser lines (typically ranging from 250 MHz to 1 GHz) by means of a series of Fabry-Perot cavities \cite{SteinmetzWilken,MurphyUdem07,BrajeKirchner08,WilkenLovis10,DoerrKentischer12,CurtoManescau12,GriestWhitmore10,OstermanDiddams07,OstermanYcas12,LiBenedick08,LiGlenday10, BenedickChang10}. 


However, observations of galaxy structures and detailed studies of the Milky Way require deployment of spectrographs in the low- and medium-resolution range. For this type of spectrograph, an OFC needs to provide stable spacings from 50~GHz to a few hundreds of GHz which is only hardly achievable in mode-locked lasers due to the laser cavity geometries. As for the semiconductor micro-ring cavities and toroids that are able to provide OFCs with spacing up to a few hundreds of GHz, they suffer from thermal effects degrading the OFC stability which makes them difficult candidates for application in astronomy \cite{ZajnulinaBoggio15}.   

In our group, a fibre-based approach for the generation of OFC suitable for spectrographs in the low- and medium resolution range has been proposed and extensively studied experimentally and by means of numerical simulations \cite{ZajnulinaBoggio15,ZajnulinaBoehm15, BoggioRieznik12, ZajnulinaBoggio13, ZajnulinaBoehm14}. Contrary to mode-locked lasers used for generation of OFCs, our approach is a single pass consisting of two fibre stages with a conventional single-mode fibre as the first stage and a suitably pumped amplifying Erbium-doped fibre with anomalous dispersion as the second stage. The initial input field is generated by two equally intense continuous-wave (CW) lasers spectrally separated by the so called laser frequency separation ($LFS$). The evolution of a OFC begins in the first fibre due to cascaded a four-wave mixing process. In the second fibre, the OFC from the first fibre stage gets broadened due to a further four-wave mixing as well as to a soliton-compression effect based on the extreme optical pulse amplification \cite{ZajnulinaBoggio15}. Our approach is simple, robust, low-cost, and versatile: it can also be deployed for OFC generation for spectroscopy applications as well as for subpicosecond pulse generation for telecommunication.

The applications in the astronomy requires broadband OFCs with sharp spectral lines. Therefore, it is crucial to precisely understand and control the build-up of the optical pulse temporal shape in the first fibre stage of our approach, because the temporal shapes will determine the bandwidth and stability of produced OFCs. Moreover, the temporal shapes of the first fibre output will directly influence the pulse build-up in the second amplifying stage. 
Further, any temporal aperiodicity of the optical pulses within the output pulse train would result in the broadening of the OFC lines and, therefore, needs to be prevented to achieve sharp OFC lines at the end. 

However, the Generalised Nonlinear Schr\"odinger Equation (GNLS) with a bichromatic initial condition that we use to model and study our approach is not integrable which hinders us from a thorough understanding of the pulse shape build-up in the first fibre stage. Thus, we apply the numerical technique called Soliton Radiation Beat Analysis (SRBA) to get insight into the pulse formation in the first fibre \cite{BoehmMitschke06, BoehmMitschke07}. This technique is capable of dealing with nonintegrable equations with arbitrary initial conditions and allows to retrieve the soliton content if the optical pulses generated in fibre-based systems.

In our previous work (\cite{ZajnulinaBoehm15}), using the SRBA and a fixed value of the laser frequency separation of $LFS = 78.125\,\mathrm{GHz}$, we identified a state of free, i.e. separated, solitons for higher laser input powers ($>$ 3~W), an intermediate state that denotes a continuous dissolving of a soliton crystal into free solitons in the moderate input-power region (1.3~W - 3~W), and a soliton crystal state for low input powers ($<$ 1.3~W) characterised by a common propagation of several optical pulses with a fixed time separation. We stated that the intermediate state is most suitable for generation of the OFCs for astronomical spectrograph calibration, because it combines the properties of a soliton crystal that guarantees stable spectral spacings of the OFC lines with simple dynamics of separated fundamental solitons.  

The discovered soliton crystal state constitutes an unusual nonlinear pattern.
  The strict temporal periodicity of its components that corresponds to the value of $LFS$ makes it an interesting object of studies
with high level of  potential in applications where fixed pulse temporal separation denoting low level of timing jitter is required.
Moreover, the soliton crystal oscillates with the fundamental frequency $Z_0$ over the propagation distance as we observed in our previous work (\cite{ZajnulinaBoehm15}).
For us, the knowledge of the dependence of the soliton crystal properties such as, for instance, its fundamental frequency on different fibre parameters
(group-velocity dispersion (GVD), nonlinearity, and optical losses) and on the initial light source characteristics ($LFS$ and laser-intensity variations) is necessary for a successful experimental realisation of our approach for generation of OFC for calibration of low- and medium resolution spectrographs.

Here, we present our studies on the soliton crystal properties with respect to the fibre parameters and the initial light source characteristics. We found out that the soliton crystal can be fully described by two quantities, namely by its fundamental frequency and the laser input-power threshold at which the dissolution of the soliton crystal into the state of free solitons takes place. Further, it exhibits features of a linear and nonlinear optical pattern at the same time. Thus, the appearance of the fundamental frequency is a purely linear effect, whereas the crystal has similar properties as separated solitons in terms of the dispersion and nonlinear length which represents its nonlinear nature. Moreover, the soliton crystal is insensitive to laser input power fluctuations which makes its experimental realisation relatively simple to be implemented. Again, we performed our studies using the SRBA technique.

This paper is structured as follows: in Sec.~\ref{sec:setup}, we present the experimental setup for generation of OFC in fibres and the corresponding mathematical model, the concept of SRBA and the interpretation of the SRBA power spectra are discussed in Sec.~\ref{sec:method},
the results of our numerical studies are presented in Sec.~\ref{sec:results}, and a conclusion is drawn in Sec.~\ref{sec:conclusion}.

\section{Experimental setup and mathematical model}
\label{sec:setup}
\begin{figure}[htp]
 \centering
 \includegraphics[width=\columnwidth]{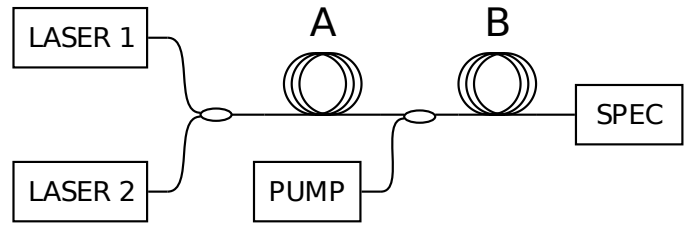}
 \caption{Schematic representation of the proposed approach for generation of optical frequency combs. LASER 1: fixed CW laser, LASER 2: tuneable CW laser, A: conventional single-mode fibre, B: Er-doped fibre, PUMP: pump laser for B, SPEC: astronomical spectrograph \cite{ZajnulinaBoehm15}}
 \label{fig:setup}
\end{figure}

Fig.~\ref{fig:setup} shows the schematic representation of the experimental setup for generation of OFC for spectral calibration of astronomical spectrographs in the low- and medium resolution range \cite{ZajnulinaBoggio15, ZajnulinaBoehm15, BoggioRieznik12, ZajnulinaBoggio13, ZajnulinaBoehm14}. In this figure, A is a conventional single-mode fibre, whereas B is a suitably pumped Erbium-doped fibre with anomalous dispersion. The generation of a comb begins with two equally intense CW lasers (Laser 1 and Laser 2). They are independent, free-running and feature relative frequency stability of $10^{-8}$ over a one-hour time frame
which is sufficient for astronomical applications in the low- and medium-resolution range.
Laser 1 has a fixed angular frequency $\omega_1$, whereas Laser 2 has frequency $\omega_2$ that is tuneable. The resulting central frequency is $\omega_{c} = (\omega_1 + \omega_2 )/2$ coinciding with the central wavelength at 1531\,nm. The laser frequency separation is given by $LFS = |\omega_1 - \omega_2|/(2\pi)$.

\begin{figure}[htp]
 \centering
 \includegraphics[width=\columnwidth]{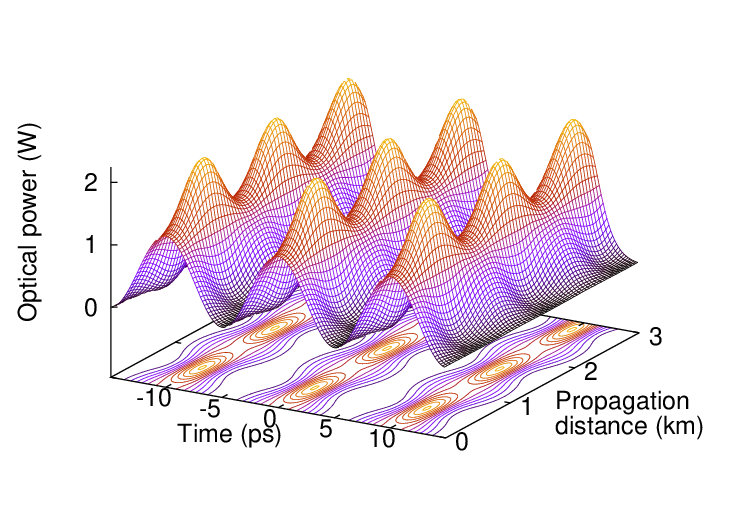}
 \caption{Spatio-temporal evolution of a soliton crystal with input power of $P_0$ = 1.3~W,
 laser frequency separation $LFS = 100\,\mathrm{GHz}$,
 group-velocity dispersion parameter $\beta_2 = - 15\,\mathrm{\frac{ps^2}{km}}$,
 and the nonlinear coefficient $\gamma = 2\,\mathrm{W^{-1}km^{-1}}$}
 \label{fig:3Dpropagation}
\end{figure}

An initial OFC arises in the first fibre due to a cascade of four-wave mixing processes,
whereas in the second fibre, the OFC is broadened due to the strong pulse amplification and simultaneous compression. Pulse compression in an amplifying fibres is known from the late 80's and is an alternative technique to the compression in dispersion-decreasing fibres \cite{BlowDoran88, ChernikovDianov93, VoroninZheltikov08, LiKutz10}.
As we discovered in our previous work (\cite{ZajnulinaBoehm15}), in fibre A, a common soliton crystal state arises out of the initial laser field for low input powers.
For high input powers, free, i.e. separated, solitons are formed.
In between, there is an intermediate state that denotes a continuous dissolving of the soliton crystal into the free solitons with increasing input power.
The soliton crystal constitutes a bound state of several optical pulses that propagate through the optical fibre with a fixted temporal separation. 
Fig. \ref{fig:3Dpropagation} shows an example of the temporal and spatial evolution of a soliton crystal.

Due to the strict temporal periodicity of crystal's pulse components, its OFC spectrum exhibits sharp spectral lines with stable spacings which makes the soliton crystal a promising nonlinear pattern for high-quality OFC generation. However, a better understanding is still needed about how such a soliton crystal arises. Here, we study the build-up of a soliton crystal in detail.
Specifically, we address the question how does the soliton crystal build-up depend on different fibre parameters (GVD, nonlinearity, and optical losses)
and the initial light characteristics ($LFS$ and laser-intensity variation).
We model the nonlinear light propagation in this fibre by means of the GNLS for a slowly varying optical field envelope $A = A(z, t)$ in the co-moving frame \cite{VoroninZheltikov08, Agrawal12, Agrawal08, TuritsynBale12}:
\begin{eqnarray} \label{eqn:GNLS}
 \frac{\partial A}{\partial z} & = & i \sum_{j=2}^{3} \frac{i^j}{j!} \beta_j \frac{\partial^j A}{\partial t^j} +
 i \gamma \left( 1 + \frac{i}{\omega_0} \frac{\partial}{\partial t} \right)\times\\
 & & \left( A \int\limits_{-\infty}^{\infty} R(t') |A(t - t')|^2 \mathrm{d} t' \right) \nonumber
 - \frac{\alpha}{2} A,
\end{eqnarray}
where $\beta_j = \left. \frac{\partial^j \beta}{\partial \omega^j}\right|_{\omega=\omega_0}$
is the value of the $j^\mathrm{th}$ dispersion order at the carrier frequency $\omega_\mathrm{c}$,
$\gamma = \frac{\omega_\mathrm{c} n_2}{c \cdot S}$ is the nonlinear coefficient with
$n_2$ being the nonlinear refractive index of silica, $S$  the effective mode area,
$c$ the speed of light, and $\alpha$ the linear optical losses due to the material absorption.
The response function $R(t)$ incorporates both,
the electronic contribution assumed to be instantaneous and the contribution from vibrational modes of silica molecules,
the delayed Raman effect is included into it via the function $h_\mathrm{R}(t)$:
\begin{equation}
 R(t) = (1 - f_\mathrm{R}) \delta(t) + f_\mathrm{R} h_\mathrm{R} (t)
\end{equation}
with $f_\mathrm{R} = 0.245$ representing the fraction of the delayed Raman response to the total nonlinear polarisation and $h_\mathrm{R}(t)$ defined as:

\begin{eqnarray}
 \displaystyle
 h_\mathrm{R}(t) & = & (1 - f_b) h_a (t) + f_b h_b (t), \\
 h_a(t) & = & \frac{\tau_1^2 + \tau_2^2}{\tau_1 \tau_2^2} \exp\left(- \frac{t}{\tau_2}\right) \sin\left(- \frac{t}{\tau_1}\right),\\
 h_b(t) & = & \left( \frac{2\tau_b - t}{\tau_b^2}\right) \exp\left(- \frac{t}{\tau_b}\right),
\end{eqnarray}
where $\tau_1 = 12.2\,\mathrm{fs}$ and $\tau_2 = 32\,\mathrm{fs}$ are the characteristic times
of the Raman response of silica and $f_b = 0.21$ is the corresponding vibrational
instability with $\tau_b \approx 96\,\mathrm{fs}$. \cite{ZajnulinaBoehm15, Agrawal12, Agrawal08}

The initial condition that represents the radiation of two CW lasers is described by \cite{ZajnulinaBoehm15, ZajnulinaBoggio15, BoggioRieznik12, ZajnulinaBoggio13, ZajnulinaBoehm14}:
\begin{equation} \label{eqn:initial}
 A(z=0, t) = N \sqrt{P_0} \cos(\omega_\mathrm{c} t)
\end{equation}
with $P_0$ being the initial laser power and $\omega_\mathrm{c}$ the central frequency. $N$ is a scale factor that plays the role of separated soliton scale order in the case of free (separated) solitons. It is determined by the relation
\begin{equation} \label{eqn:N}
 N^2 = \frac{L_\mathrm{D}}{L_\mathrm{NL}} = \frac{\gamma P_0}{(2\pi\cdot LFS)^2 |\beta_2|} ,
\end{equation}
where $L_\mathrm{D} = \frac{1}{(2\pi\cdot LFS)^2 |\beta_2|}$ is the dispersion length and
$L_\mathrm{NL} = \frac{1}{\gamma P_0}$ the nonlinear length \cite{ZajnulinaBoehm15, ZajnulinaBoehm14, Agrawal12, Agrawal08, TuritsynBale12}.
In the initial condition Eq.~\ref{eqn:initial}, $N$ equals 1. The dispersion length $L_\mathrm{D}$ relates to the GVD presented via the second order dispersion $\beta_2$ in Eq.~\ref{eqn:GNLS} and is defined as the distance in which an optical pulse has broadened to $\sqrt 2$ its initial width during the propagation through the fibre \cite{Mitschke10}.
The nonlinear length $L_\mathrm{NL}$ is the distance in which the pulse undergoes a nonlinear phase rotation of $\pi$ at its maximum \cite{TuritsynBale12}. This length corresponds to the self-phase modulation that is the first effect in the second term on the right-hand side of Eq.~\ref{eqn:GNLS}. One of the goals of this paper is to investigate if soliton crystals have a similar behaviour as separated solitons in terms of GVD and nonlinearity and can also be described by dispersion and nonlinear lengths.

The numerical integration of Eq.~\ref{eqn:GNLS} and Eq.~\ref{eqn:initial} is performed using the fourth-order Runge-Kutta in the interaction picture method in combination with the local error method within a temporal window of 128\,ps sampled with 214 points \cite{Hunt07}.

Note that the analytic solution of the GNLS that describes a soliton crystal
can be constructed using a very tedious calculation based on a Darboux transformation as the Ref.~\cite{GuoHao13} suggests.
This is, however, only possible for an integrable form of the GNLS,
i.e. for an equation that does not contain the third-order dispersion, the Raman effect, and the optical losses.
Especially, the Raman effect expressed as a retarded integral and the optical losses that
we include into consideration here destroy the integrability of our GNLS (Eq.~\ref{eqn:GNLS}).
This is why we use the numerical technique of the SRBA to get a deeper insight into the soliton crystal properties.

\section{Method: soliton radiation beat analysis}
\label{sec:method}
As the optical field propagates through the fibre, the peak power oscillates over the propagation distance. These oscillations contain information about the solitons involved since they come around due to the beating between the individual solitons or the beating between the solitons and the dispersive-waves background. The content of these solitons is retrieved by means of the SRBA as follows: first, the optical field along the propagation distance, i.e.
$A(z, t)$, is calculated for a given value of the input power $P_0$, then the optical power $P(z, t) = |A(z, t)|^2$ is determined. After that, the optical power $P(z)$ is extracted at the pulse centre at $t = 0$, i.e. $P(z) = P(z, t = 0)$, and the power spectrum $\tilde{P}(Z)$ is then obtained via a Fourier transform of $P(z)$ with respect to the spatial coordinate $z$.
The coordinate $Z$ has the meaning of the spatial frequency in Fourier space \cite{ZajnulinaBoehm15}. The assembling of the $\tilde{P}(Z)$ data into a surface plot for different input power values yields a typical SRBA power spectrum graph presenting the oscillation frequencies that occur if solitons (provided that any solitons are involved) beat with each other or if the beating between a single soliton and the dispersive-waves background takes place. To increase the visibility of such oscillation frequencies in an SRBA graph,
we apodise the optical power $P(z)$ by means of an apodisation function $f(k)$
before Fourier transforming it into the spatial frequency domain \cite{BoehmMitschke06, BoehmMitschke07}:
\begin{equation}
 f(k) = \exp\left[-\left(\frac{k-K/2}{bK}\right)^2\right]
\end{equation}
with $1/b$ being the apodisation strength and $k \in [1, \dots , K]$, where $K = 50000$ is the total number of distance sampling points. For our studies, we set $b = 0.2$ \cite{ZajnulinaBoehm15}. Further, the resolution of the oscillation frequencies in an SRBA power spectrum plot strongly depends on the total fibre length chosen for simulation.
Precisely, it scales as $1/L$ with $L$ being the total propagation length \cite{ZajnulinaBoehm15}. Therefore, in the course of our studies, we set the fibre length to $L = 50\,\mathrm{km}$ to obtain well-resolved, i.e. sharp, frequency lines.
Moreover, the input power steps for our simulations are set to $\Delta P_0 = 0.02\,\mathrm{W}$ which will also guarantee a high level of the power spectrum resolution.

\begin{figure}[htp]
 \centering
 \includegraphics[width=\columnwidth]{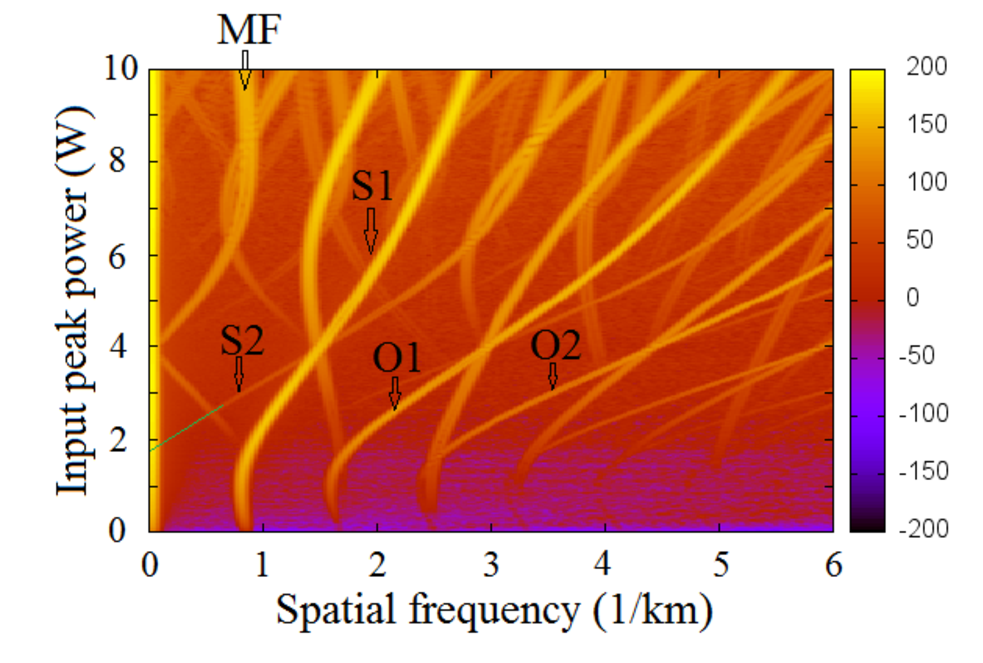}
 \caption{Power spectrum for different values of input power $P_0$,
 laser frequency separation $LFS = 100\,\mathrm{GHz}$,
 group-velocity dispersion parameter $\beta_2 = - 15\,\mathrm{\frac{ps^2}{km}}$,
 and the nonlinear coefficient $\gamma = 2\,\mathrm{W^{-1}km^{-1}}$
}
 \label{fig:SRBA2}
\end{figure}

To show how to interpret an SRBA power spectrum,
we choose a spectrum for the input powers $0\,\mathrm{W} < P_0 \le 10\,\mathrm{W}$ (Fig.~\ref{fig:SRBA2}). This spectrum was obtained for $LFS = 100\,\mathrm{GHz}$, the GVD parameter $\beta_2 = - 15\,\mathrm{\frac{ps^2}{km}}$, $\beta_3 = - 0.1\,\mathrm{\frac{ps^3}{km}}$, and the nonlinear coefficient $\gamma = 2\,\mathrm{W^{-1}km^{-1}}$. The optical losses are set to 0, i.e. $\alpha = 0\,\mathrm{\frac{dB}{km}}$.

First, the visibility of a spatial frequency in an SRBA graph is a measure of the intensity of the involved components, i.e. the brighter or the more visible a beating frequency is, the higher is the amplitude of the components that constitute this frequency.

Second, one will see a strong peak for any values of $P_0$ at the spatial frequency $Z = 0\,\mathrm{km}^{-1}$ in all the SRBA graphs presented below. This peak arises during the Fourier transform from the optical into the frequency domain and corresponds to the average value of the optical power $P(z)$. Since the origin of this peak has a purely mathematical nature, it does not contain any information about the nonlinear soliton dynamics. Therefore, we will exclude it from consideration.

Third, single solitons arise at a positive threshold value of the input power, i.e. for $P_0 > 0\,\mathrm{W}$, and evolve depending on $\sqrt{P_0}$ \cite{Taylor92}. Accordingly, the beating frequency of a single soliton with other solitons or the beating of a soliton with a dispersive-waves background has an input-power threshold in an SRBA power spectrum. In Fig.~\ref{fig:SRBA2}, the beating between a single soliton and the dispersive-waves background that is represented by the branch S2 arises at $P_0 = 1.8\,\mathrm{W}$ and $Z = 0\,\mathrm{km}^{-1}$.
According to Eq.~\ref{eqn:N}, the scale order of the corresponding soliton is $N = 0.78$.
In the case of a Nonlinear Schr\"odinger Equation without the additional terms describing the higher-order dispersion, the shock, the Raman effect, and the optical losses, fundamental solitons are created for $0.5 \le N < 1.5$ \cite{Taylor92}. Since the scale order of the S2 soliton lies in this interval, we can identify it as a fundamental soliton.

Fourth, the most prominent soliton branch S1 starts at the spatial frequency $Z_0 = Z(P_0 = 0\,\mathrm{W}) = 0.84\,\mathrm{km}^{-1}$. For input powers of approximately $P_0 > 2 W$, this branch has a similar behaviour to the branch S2 and, thus, can be identified as a beating between a single soliton and the dispersive-waves background. For input powers $P_0 \rightarrow 0\,\mathrm{W}$, the behaviour is, however, different: the spatial frequency changes only insignificantly with the value of the input power. This is the behaviour of a soliton crystal state that is formed, because the cosine-function incorporated into the initial condition Eq. \ref{eqn:initial} provides an infinite amount of energy for $t \to \pm \infty$ that is sufficient to create such a collective state even for very low input powers,
i.e. for $P_0 \to 0\,\mathrm{W}$. The spatial frequency $Z_0$ is the fundamental feature of a soliton crystal. The soliton crystal dissolves continuously into the state of free solitons as the input power increases. At least, there is no visible transition between the soliton crystal and the state of free solitons (cf. Ref.~\cite{ZajnulinaBoehm15}).

Fifth, the relative energy content of the different components can be determined by comparing the frequencies of the branches. Thus, for higher input powers, S1 has less energy than S2 and, therefore, oscillates slower. So, S1 has the spatial frequency $Z = 2.8\,\mathrm{km}^{-1}$ and S2 the frequency $Z = 3.6\,\mathrm{km}^{-1}$ at $P_0 = 10\,\mathrm{W}$.

Sixth, the branches O1, O2 in Fig.~\ref{fig:SRBA2} denote the overtones of the branch S2. Since the overtones provide no addition information about the soliton content, they will be also excluded from further consideration in this study.

Seventh, the mixing frequencies are given as integral multiples of the difference and sum frequencies of different branches. Thus, MF constitutes a difference frequency of branches S1 and S2 (Fig.~\ref{fig:SRBA2}). Since the mixing frequencies do not give us further information about the content of the solitons involved, we will exclude them from  consideration\cite{ZajnulinaBoehm15, BoehmMitschke06, BoehmMitschke07}.

\section{Results}
\label{sec:results}
Having introduced what is the SRBA technique and how to interpret the SRBA power spectra,
we proceed with the presentation of the results of our studies on the soliton crystal properties depending on the GVD parameter $\beta_2$,
the laser frequency separation $LFS$, the nonlinear coefficient $\gamma$, the optical losses $\alpha$, and the difference in the laser intensity.

\subsection{Fundamental frequency $Z_0$ depending on the laser frequency LFS and the group-velocity dispersion GVD}
\begin{figure}[htp]
 \centering
 \includegraphics[width=\columnwidth]{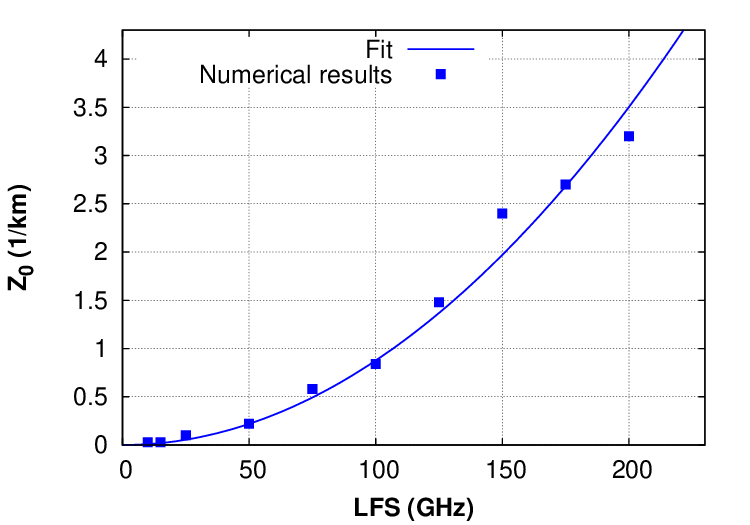}
 \caption{Fundamental frequency $Z_0$ depending on the laser frequency separation $LFS$}
 \label{fig:SRBA3}
\end{figure}

A soliton crystal is characterised by its fundamental frequency at which it arises for input power value 0, i.e. $Z_0 = Z(P_0 = 0\,\mathrm{W})$ (Sec.~\ref{sec:method}).
Fig.~\ref{fig:SRBA3} shows how this frequency depends on different values of the laser frequency separation $LFS.$ To obtain this graph, the following fibre parameters were chosen:
$\beta_2 = - 15\,\mathrm{\frac{ps^2}{km}}$, $\gamma = 2\,\mathrm{W^{-1}} \mathrm{km^{-1}}$, and $\alpha = 0\,\mathrm{\frac{dB}{km}}$. The laser frequency separation took the values in the range $10\,\mathrm{GHz} \le LFS \le 200\,\mathrm{GHz}$. As one can see, the dependence of $Z_0$ on $LFS$ is quadratic. This can be seen from the excellent fit to the parabolic curve plotted in Fig.~\ref{fig:SRBA3}.

\begin{figure}[htp]
 \centering
 \includegraphics[width=\columnwidth]{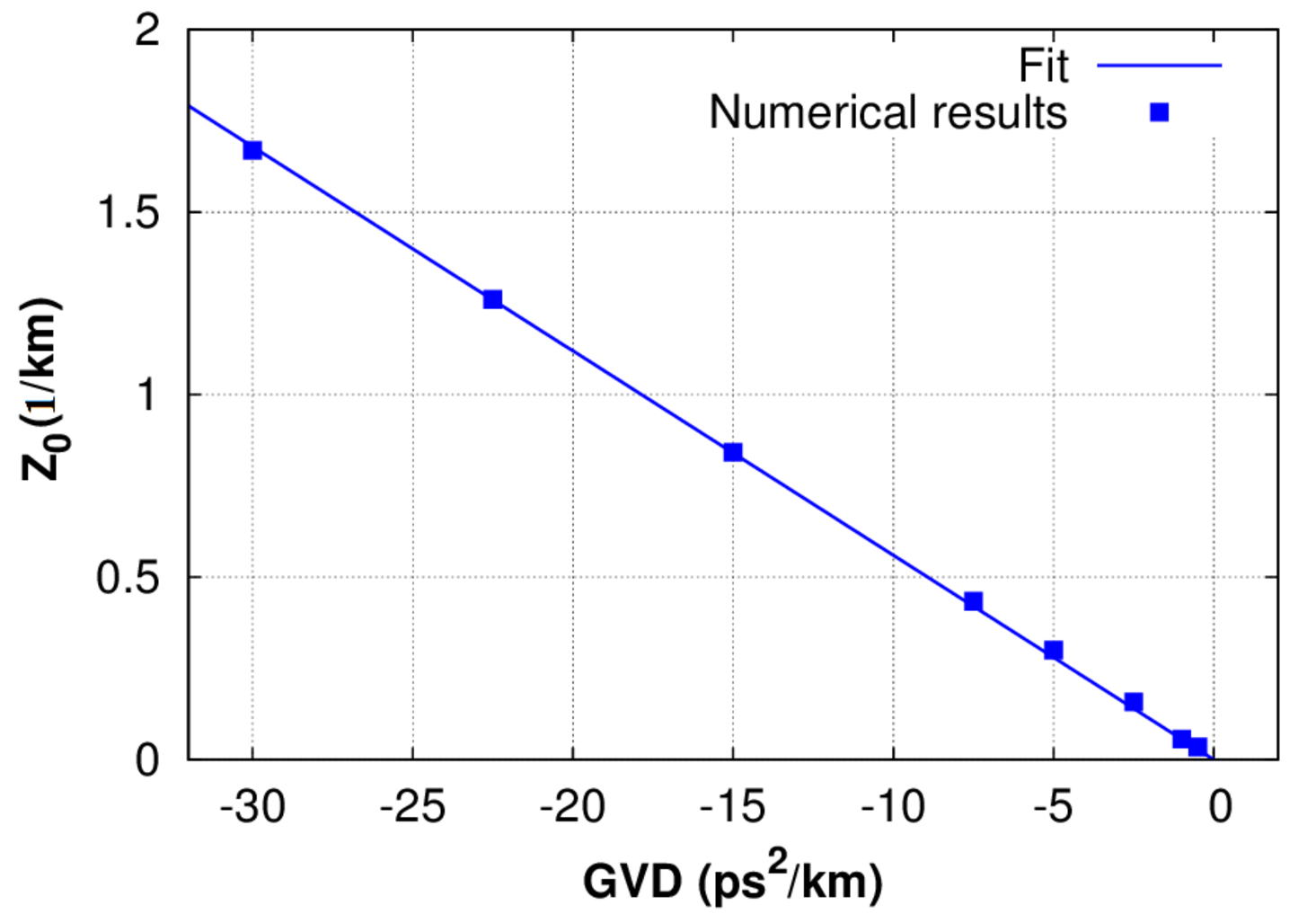}
 \caption{Fundamental frequency $Z_0$ depending on the GVD parameter $\beta_2$}
 \label{fig:SRBA4}
\end{figure}

In Fig.~\ref{fig:SRBA4}, the laser frequency separation was fixed to $LFS = 100\,\mathrm{GHz}$, whereas the GVD parameter was varied from $\beta_2 = -30\,\mathrm{\frac{ps^2}{km}}$ to $\beta_2 = -0.5\,\mathrm{\frac{ps^2}{km}}$
All other system parameters were the same as previously. The dependence of $Z_0$ on $\beta_2$ is clearly linear.

The results presented here give rise to the conclusion that the dispersion length of a soliton crystal $L_\mathrm{D}^\mathrm{SC}$ obeys the same relationship as the dispersion length of a single (separated) soliton, namely
\begin{equation}
L_\mathrm{D}^\mathrm{SC} = X \frac{1}{(2\pi\cdot LFS)^2 |\beta_2|}
\end{equation}
with $X$ being a conversion factor with the value of $X \approx 2\pi$.

\subsection{Impact of the fibre nonlinearity}
\begin{figure}[htp]
 \centering
 \includegraphics[width=\columnwidth]{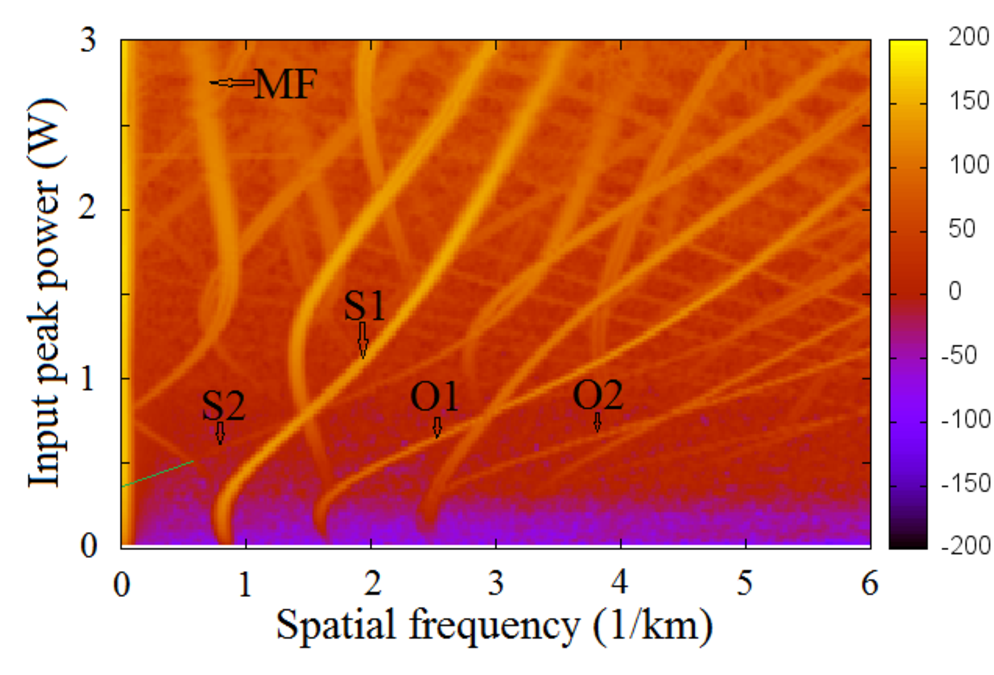}
 \caption{ Power spectrum for different values of input power $P_0$,
           laser frequency separation $LFS = 100\,\mathrm{GHz}$,
           group-velocity dispersion parameter $\beta_2 = -15\,\mathrm{\frac{ps^2}{km}}$,
           and the nonlinear coefficient $\gamma = 10 \,\mathrm{W^{-1}} \mathrm{km^{-1}}$
}
 \label{fig:SRBA5}
\end{figure}

Fig. \ref{fig:SRBA5} shows an SRBA power spectrum for the nonlinear coefficient
$\gamma = 10 \,\mathrm{W^{-1}} \mathrm{km^{-1}}$. Other fibre parameters are:
$\beta_2 = -15\,\mathrm{\frac{ps^2}{km}}$ and $\alpha = 0\,\mathrm{\frac{dB}{km}}$,
the laser frequency separation is $LFS = 100\,\mathrm{GHz}$.

If we compare Fig.~\ref{fig:SRBA5} with Fig.~\ref{fig:SRBA2}, we see that the main structures have almost the same characteristics. Thus, the fundamental frequency of the branch S1 has the same value of $Z_0 = 0.84\,\mathrm{km}^{-1}$ in both graphs.
Also fundamental frequencies of the overtones O1 and O2 coincide in both SRBA spectra.
We obtained the same results -- not presented here -- using other values of the nonlinear coefficient, namely $\gamma = 1\,\mathrm{W^{-1}} \mathrm{km^{-1}}$, $\gamma = 4\,\mathrm{W^{-1}} \mathrm{km^{-1}}$, and $\gamma = 6\,\mathrm{W^{-1}} \mathrm{km^{-1}}$.
This means that the fundamental frequency of a soliton crystal $(Z_{0})$
does not depend on the fibre nonlinearity. Its appearance is rather a linear effect that is intrinsic to the cosine function via which the initial condition is expressed (Eq.~\ref{eqn:initial}).

The main difference between the graphs is that the significant changes in the frequencies' behaviour occur at lower input powers if the nonlinear coefficient value is high. For instance, the input power threshold at which the free soliton branch S2 arises is $P_0 = 0.36\,\mathrm{W}$ for $\gamma = 10\,\mathrm{W^{-1}} \mathrm{km^{-1}}$ (it was $P_0 = 1.8\,\mathrm{W}$ for $\gamma = 2\,\mathrm{W^{-1}} \mathrm{km^{-1}}$). The mixing frequency MF cuts the axis at $Z = 0\,\mathrm{km}^{-1}$ for $P_0 = 0.76\,\mathrm{W}$ and $\gamma = 10\,\mathrm{W^{-1}} \mathrm{km^{-1}}$ and for $P_0 = 3.8\,\mathrm{W}$ and $\gamma = 2\,\mathrm{W^{-1}} \mathrm{km^{-1}}$. These results lead to the conclusion that the input-power axis scales linearly with the value of the nonlinear coefficient $\gamma$.
The evolution of the branch S1 follows this scaling. In fact, the regime of free solitons occurs for $P_0 > 1\,\mathrm{W}$ for $\gamma = 10\,\mathrm{W^{-1}} \mathrm{km^{-1}}$,
whereas it occurred at approximately $P_0 > 2\,\mathrm{W}$ for $\gamma = 2\,\mathrm{W^{-1}} \mathrm{km^{-1}}$ (cf. Sec.~\ref{sec:method}). For $P_0 \le 1\,\mathrm{W}$, the evolution of S1 is the same for $\gamma = 10\,\mathrm{W^{-1}} \mathrm{km^{-1}}$ as in the case of $\gamma = 2\,\mathrm{W^{-1}} \mathrm{km^{-1}}$ and $P_0 \le 2.5\,\mathrm{W}$.
Thus, we can draw a conclusion that the nonlinear length of a soliton crystal $L_\mathrm{NL}^\mathrm{SC}$ has the same relationship as the nonlinear length of a single soliton, i.e.
\begin{equation}
 L_\mathrm{NL}^\mathrm{SC} = \frac{1}{\gamma P_0}.
\end{equation}

\subsection{Impact of the optical losses}
\begin{figure}[htp]
 \centering
 \includegraphics[width=\columnwidth]{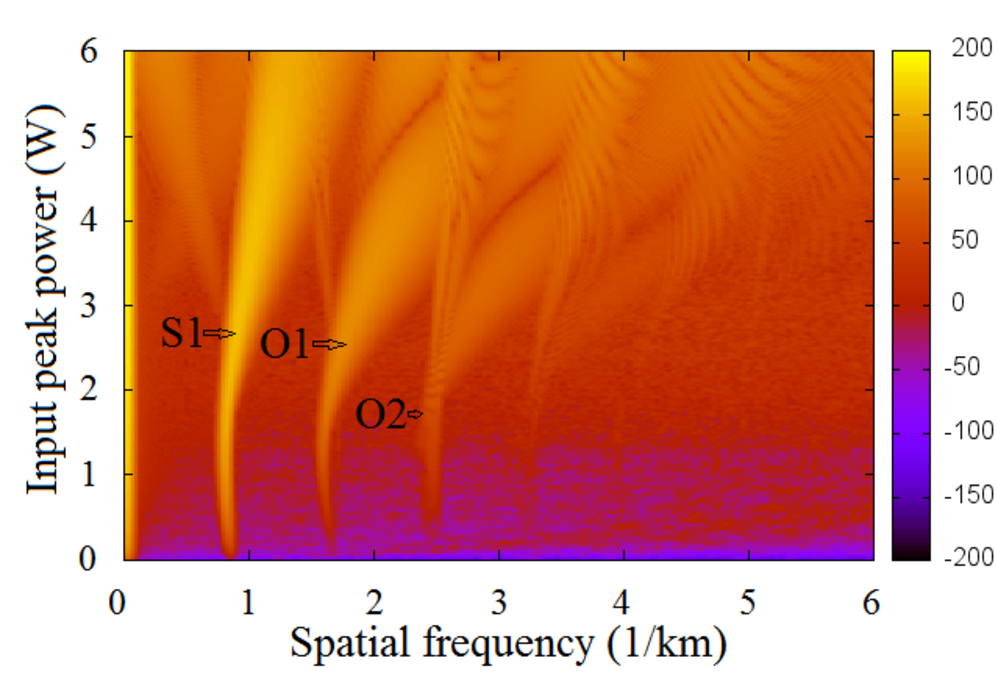}
 \caption{Power spectrum for different values of input power $P_0$,
          laser frequency separation  $LFS = 100\,\mathrm{GHz}$,
          group-velocity dispersion parameter $\beta_2 = -15\,\mathrm{\frac{ps^2}{km}}$,
          the nonlinear coefficient $\gamma = 2 \,\mathrm{W^{-1}} \mathrm{km^{-1}}$,
          and optical losses $\alpha = 0.05\,\mathrm{\frac{dB}{km}}$
          }
 \label{fig:SRBA6}
\end{figure}

\begin{figure}[htp]
 \centering
 \includegraphics[width=\columnwidth]{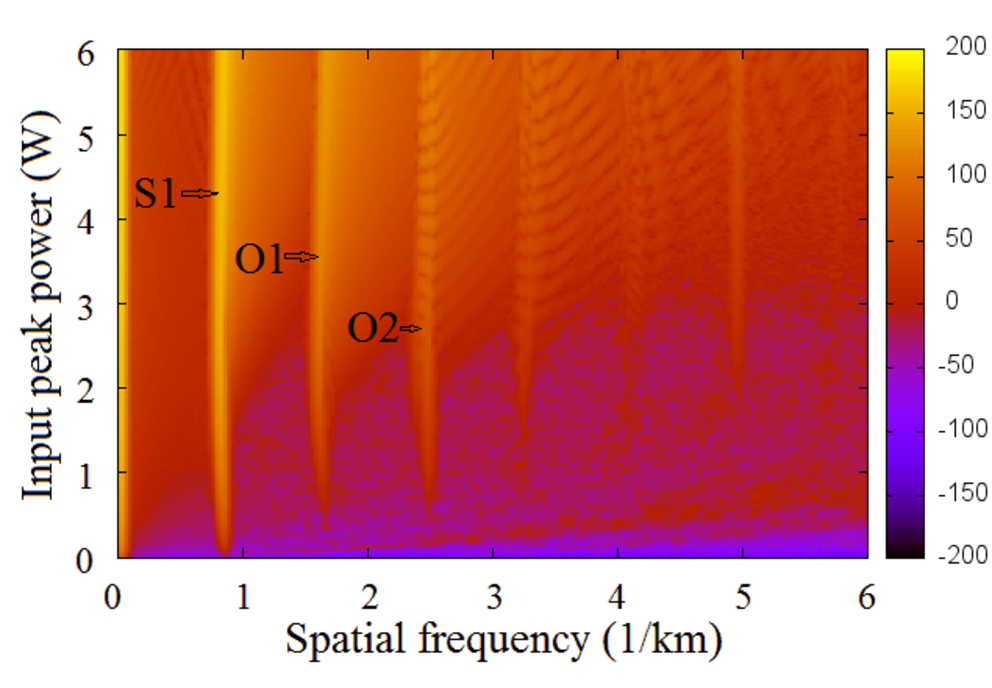}
 \caption{Power spectrum for different values of input power $P_0$,
          laser frequency separation  $LFS = 100\,\mathrm{GHz}$,
          group-velocity dispersion parameter $\beta_2 = -15\,\mathrm{\frac{ps^2}{km}}$,
          the nonlinear coefficient $\gamma = 2 \,\mathrm{W^{-1}} \mathrm{km^{-1}}$,
          and optical losses $\alpha = 0.2\,\mathrm{\frac{dB}{km}}$
          }
 \label{fig:SRBA7}
\end{figure}

We have so far considered cases in which the linear optical losses were neglected, i.e. $\alpha = 0\,\mathrm{\frac{dB}{km}}$. Now, we analyse the impact of the losses on the evolution of the soliton crystal. For that, we consider SRBA spectra for $\alpha = 0.05\,\mathrm{\frac{dB}{km}}$ (Fig.~\ref{fig:SRBA6}) and $\alpha = 0.2\,\mathrm{\frac{dB}{km}}$ (Fig.~\ref{fig:SRBA7}). Whereas the optical losses of $\alpha = 0.05\,\mathrm{\frac{dB}{km}}$ still constitute an ideal case similar to one with
$\alpha = 0\,\mathrm{\frac{dB}{km}}$, the losses of $\alpha = 0.2\,\mathrm{\frac{dB}{km}}$ are typical for the conventional single-mode fibres at wavelengths around $1.55\,\mathrm{\mu m}$ \cite{VoroninZheltikov08, LiKutz10}. To obtain the SRBA power spectrum graphs, the following system parameters were chosen: $\beta_2 = -15\,\mathrm{\frac{ps^2}{km}}$, $\gamma = 2\,\mathrm{W^{-1}} \mathrm{km^{-1}}$, and $LFS = 100\,\mathrm{GHz}$.

If we compare Fig.~\ref{fig:SRBA6} and Fig.~\ref{fig:SRBA7} with Fig.~\ref{fig:SRBA2},
we see that there is no appearance of the single soliton branch S2 when the optical losses are present. That means that, due to the optical losses, the system does not have enough energy to create a single soliton in addition to the soliton crystal branch S1. As for S1 itself, it still arises at the fundamental frequency $Z_0 = 0.84\,\mathrm{km^{-1}}$
for both values, $\alpha = 0.2\,\mathrm{\frac{dB}{km}}$ and $\alpha = 0.05\,\mathrm{\frac{dB}{km}}$ meaning that the fundamental frequency does not depend on the initial energy content of the system and its loss due to the absorption.
However, at the input power of $P_0 = 1.8\,\mathrm{W}$, the branch S2 \textquoteleft fans out\textquoteright for the loss of $\alpha = 0.05\,\mathrm{\frac{dB}{km}}$. This fan contains a continuum of spatial frequencies which can be explained as follows: the spatial frequency of free (separated) solitons scales with their energy E \cite{BoehmMitschke07b}:
\begin{equation}
 Z = \frac{E^2}{16 \pi} \frac{\gamma^2}{|\beta_2|},
\end{equation}
whereas $E$ decreases continuously with the propagation distance: $E(z) = E_0 \exp(−\alpha z)$ with $E_0$ being the initial energy of a soliton. This continuous decrease of the soliton energy has the appearance of a spatial frequency continuum as a result.

In Sec.~\ref{sec:method}, we stated that the soliton crystal dissolves continuously
into the state of separated solitons with the input power when the optical losses are neglected. Due to this continuity, we were not able to provide information on the input-power
at which the soliton crystal is dissolved into free solitons, we could only make statements about the behaviour tendencies for higher powers, i.e. $P_0 \to 6\,\mathrm{W}$, and lower input powers, i.e. $P_0 \to 0\,\mathrm{W}$. Since the appearance of a spatial frequency continuum is typical for separated solitons if there are any absorption losses present, but the soliton crystal itself (i.e. its existence and the fundamental frequency) does not depend on the optical losses, we can mark the input-power ($P_0 \to 1.8\,\mathrm{W}$) at which the spatial frequency fan begins to evolve as a point at which dissolution of a soliton crystal into free (separated) solitons occurs. Note that, if the optical losses are neglected, the emergence of the free soliton branch is also observed at $P_0 \to 1.8\,\mathrm{W}$ (Fig.~\ref{fig:SRBA2}) which supports the idea that the free soliton generation
starts at this specific $P_0$-value (threshold) for the chosen set of parameters.
According to the results presented in Sec.~\ref{sec:method}, this input power threshold depends on the fibre nonlinearity. Due to the presence of a clear threshold value, the intermediate state that features the properties of both, the soliton crystal and free solitons, is only possible in a very close vicinity of this input-power point and not in a larger
$P_0$-interval as stated in Ref.~\cite{ZajnulinaBoehm15} and Sec.~\ref{sec:intro} of this paper.

As the value of optical losses increases, the continuous fan of spatial frequencies disappears: the spatial frequencies of branches S1, O1, and O2 reach a limit that is predefined by the fundamental frequency $Z_0$ (Fig.~\ref{fig:SRBA7}). Also for higher values of $\alpha$ ($\alpha = 0.4\,\mathrm{\frac{dB}{km}}$, $\alpha = 0.8\,\mathrm{\frac{dB}{km}}$)
-- the results are omitted here -- the behaviour of S1, O1, and O2 is very similar to the case when $\alpha = 0.2\,\mathrm{\frac{dB}{km}}$. That means that, if any (realistic) optical losses are present, the pulse build-up is governed by the system characteristics at low input powers: the soliton crystal extends its range of existence to any input-powers considered,
it exists even for $P_0 = 6\,\mathrm{W}$ and $\alpha = 0.2\,\mathrm{\frac{dB}{km}}$ (Fig.~\ref{fig:SRBA7}).

\subsection{Impact of the laser intensity difference}
Since it is impossible to stabilise the laser intensities such that they are exactly equal over the duration of the calibration time (some statistical fluctuation will always occur even if the lasers exhibit a very high level of stability), we need to check what impact the difference of laser intensities has on the build-up of the soliton crystal. For that, we change the initial condition (Eq.~\ref{eqn:initial}) so that the lasers have a spectral intensity difference of 10\,\%. We consider the soliton crystal build-up in two cases, for
$\alpha = 0\,\mathrm{\frac{dB}{km}}$ and for $\alpha = 0.2\,\mathrm{\frac{dB}{km}}$. All other system parameters are the same as previously used: $\beta_2 = -15\,\mathrm{\frac{ps^2}{km}}$, $\gamma = 2\,\mathrm{W^{-1}} \mathrm{km^{-1}}$, and $LFS = 100\,\mathrm{GHz}$.
\begin{figure}[htp]
 \centering
 \includegraphics[width=0.75\columnwidth]{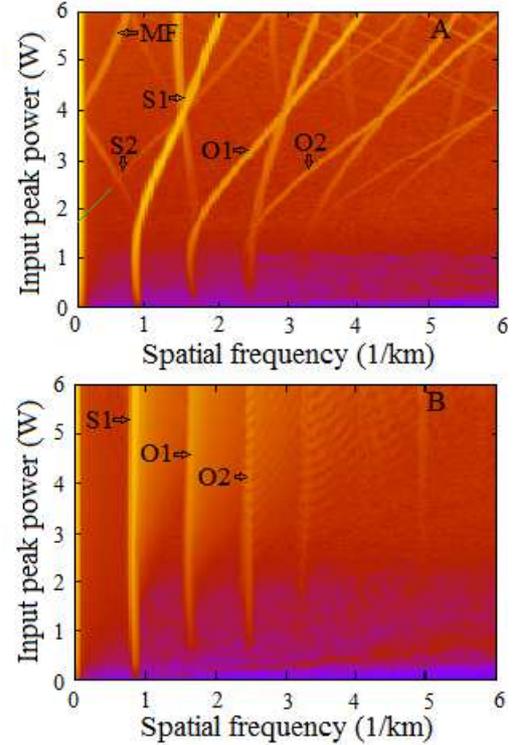}
 \caption{Power spectrum for different values of input power $P_0$,
          laser frequency separation  $LFS = 100\,\mathrm{GHz}$,
          group-velocity dispersion parameter $\beta_2 = -15\,\mathrm{\frac{ps^2}{km}}$,
          the nonlinear coefficient $\gamma = 2 \,\mathrm{W^{-1}} \mathrm{km^{-1}}$,
          and optical losses $\alpha = 0\,\mathrm{\frac{dB}{km}}$ (A) $\alpha = 0.2\,\mathrm{\frac{dB}{km}}$ (B)
          }
 \label{fig:SRBA8}
\end{figure}

The SRBA spectra obtained when the spectral laser intensities differed in the value of 10\,\% are presented in Fig.~\ref{fig:SRBA8}. If we compare Fig.~\ref{fig:SRBA8}(A) with Fig.~\ref{fig:SRBA2} and Fig.~\ref{fig:SRBA8}(B) with Fig.~\ref{fig:SRBA7}, we do not see any significant differences: the fundamental frequencies of the soliton crystal and the overtones as well as the evolution of different types of spatial frequencies with the input power $P_0$ are the same. So, provided the laser intensity difference is lower than 10\,\%, we will see no impact on the soliton crystal build-up and the evolution of free solitons, any intensity fluctuations that are smaller than 10\,\% will therefore be of no importance within the spectrograph calibration process, i.e. the laser intensity stabilisation is not a critical point in order to generate well-behaving pulses with high-quality OFC.

\section{Conclusion}
\label{sec:conclusion}
A fibre-based approach for the generation of optical frequency combs for spectral calibration of astronomical spectrographs in the low- and medium-resolution was proposed and studied in our group. This approach consists of two concatenated fibres and uses two continuous-wave lasers in the near infrared as a light source. The first fibre is a conventional single-mode fibre, the second one is a suitably pumped Erbium-doped fibre. It is crucial to precisely understand the pulse build-up in the first fibre, since the pulse shape after the first fibre will influence the pulse build-up in the second amplifying fibre. 

To get deep insight into the pulse build-up in the first fibre, we used the numerical technique of the Soliton Radiation Beat Analysis. In the previous work (\cite{ZajnulinaBoehm15}), we discovered that, for low input powers, a common soliton crystal state characterised by a bound propagation of several optical pulses with a fixed temporal separation arises out of the initial deeply modulated cosine-wave that represents the radiation of both continuous-wave lasers. As the input power increases, the soliton crystal continuously dissolves into free (separated) solitons. Here, we proceeded with the Soliton Radiation Beat Analysis of the pulse build-up in the first fibre and worked out the properties of the soliton crystal with respect to different fibre parameters (group-velocity dispersion, nonlinearity, and optical losses) and the light source characteristics (laser frequency separation and laser intensity difference).

In the course of our studies, we have identified two quantities that describe a soliton crystal. The first one is the fundamental spatial frequency, i.e.\/ the frequency at the zero input power value ($Z_0 = Z(P_0 = 0\,\mathrm{W})$), at which a soliton crystal arises. The second quantity is the input power threshold at which the soliton crystal dissolves into free (separated) solitons. The appearance of the fundamental frequency is a purely linear effect that is intrinsic to the cosine-function via which the initial condition represents the radiation of two continuous-wave lasers. Accordingly, the fundamental frequency value is not dependent on the fibre nonlinearity. The input power threshold for the dissolution of the soliton crystal into free solitons depends, however, on the fibre nonlinear coefficient: for higher values of the nonlinear coefficient, the dissolution occurs at lower input powers. In fact, the evolution of the soliton crystal and the subsequent free solitons scales linearly with the nonlinear coefficient $\gamma$ which leads to the conclusion that the nonlinear length of the soliton crystal is similar to the definition of the nonlinear length of separated solitons: $L_\mathrm{NL}^\mathrm{SC} = \frac{1}{\gamma P_0}.$

Further, the fundamental frequency depends linearly on the absolute value of the fibre group-velocity dispersion parameter $\beta_2$ and quadratically on the initial laser frequency separation $LSF$. The dispersion length obeys, accordingly, the relationship that is also typical for free solitons, namely $L_\mathrm{D}^{\mathrm{SC}} = X \frac{1}{(2\pi\cdot LFS)^2 |\beta_2|}$ with $\beta_2$ being the group-velocity dispersion parameter and $X$ a conversion factor of $X \approx 2\pi$. Our results show that the fibre parameters $\beta_2$ and $\gamma$ can be changed over a broad range without a significant impact on the soliton crystal behaviour.

The value of the soliton-crystal fundamental frequency does not depend on the optical losses. If weak optical losses of $\alpha = 0.05\,\mathrm{\frac{dB}{km}}$ are included into the analysis, the input power threshold at which the soliton crystal dissolves into free  solitonsis characterised by the appearance of a spatial frequency continuum that is typical for free solitons. At higher and more realistic values of the optical losses (for example, $\alpha = 0.2\,\mathrm{\frac{dB}{km}}$), the soliton crystal extends its range of existence to the full range of input powers considered, i.e. to $0\,\mathrm{W} < P_0 < 6\,\mathrm{W}$. That means that, if the optical fibre losses are not prevented, e.g.,
by additional fibre pumping as described in Ref.~\cite{CastanonEllingham06}, the soliton crystal will always be present in the experimental realisation. Since the soliton crystal has a simpler dynamics as compared to a train of free solitons that may be subjected to the soliton fission resulting in the appearance of additional frequency components in the comb if the fibre nonlinearity or the input power is too high, the presence of a soliton crystal for a wide range of input powers makes the proposed approach less critical to the unwanted effect of the pulse break-up, the resulting optical frequency combs are well-behaved.

Moreover, our studies showed that neither the fundamental frequency of the soliton crystal nor the input power threshold for the dissolution of the crystal into free solitons are sensitive to the laser intensity difference. Here, we chose a difference in the intensities of 10\,\% and have not observed any impact on the characteristics of the soliton crystal and the free solitons in both cases, when the optical losses of $\alpha = 0.2\,\mathrm{\frac{dB}{km}}$ are present and when they are set to zero. This makes the proposed approach robust in respect to the laser intensity fluctuations of less than 10\,\%. The fluctuation of the laser frequencies will, however, lead to the appearance of new frequency components in the optical frequency comb, whereas a high level of the optical pulse timing jitter will result in the comb line broadening. The state-of-the-art diode lasers provide, however, a sufficient level of the intensity and frequency stability to be effectively deployed for calibration of astronomical spectrographs \cite{ZajnulinaBoehm15}.

\section{Acknowledgement}
\label{sec:acknowledgement}
We acknowledge the financial support of the German Federal Ministry of Education and Research (Grant 03ZAN11).



\bibliographystyle{elsarticle-num} 


\end{document}